\documentclass{aastex63}
\usepackage{CJKutf8}
\usepackage{amsmath}

\received{Nov 14, 2022}
\revised{Dec 20, 2022}
\accepted{Dec 30, 2022}

\submitjournal{\apjl}

\shorttitle{Eccentric IHZ}
\shortauthors{}

\begin{document}
\begin{CJK*}{UTF8}{gbsn}

\title{Inner Habitable Zone Boundary for Eccentric Exoplanets }

\author[0000-0002-1592-7832]{Xuan Ji(纪璇)}
\affiliation{Department of the Geophysical Sciences, The University of Chicago, Chicago, IL 60637 USA}

\author[0000-0001-7509-0563]{Nora Bailey}
\affiliation{Department of Astronomy \& Astrophysics, The University of Chicago, Chicago, IL 60637}

\author[0000-0003-3750-0183]{Daniel Fabrycky}
\affiliation{Department of Astronomy \& Astrophysics, The University of Chicago, Chicago, IL 60637}

\author[0000-0002-1426-1186]{Edwin S. Kite}
\affiliation{Department of the Geophysical Sciences, The University of Chicago, Chicago, IL 60637 USA}

\author[0000-0002-5929-8951]{Jonathan H. Jiang}
\affiliation{Jet Propulsion Laboratory, California Institute of Technology, Pasadena, CA 91109}

\author{Dorian S. Abbot}
\affiliation{Department of the Geophysical Sciences, The University of Chicago, Chicago, IL 60637 USA}

\begin{abstract}

The climate of a planet can be strongly affected by its eccentricity due to variations in the stellar flux. There are two limits for the dependence of the inner habitable zone boundary (IHZ) on eccentricity: (1) the mean-stellar flux approximation ($S_{\mbox{IHZ}} \propto \sqrt{1-e^2}$), in which the temperature is approximately constant throughout the orbit, and (2) the maximum-stellar flux approximation ($S_{\mbox{IHZ}} \propto (1-e)^2$), in which the temperature adjusts instantaneously to the stellar flux. Which limit is appropriate is determined by the dimensionless parameter $\Pi = \frac{C}{BP}$, where $C$ is the heat capacity of the planet, $P$ is the orbital period, and $B=\frac{\partial \Omega}{\partial T_s}$, where $\Omega$ is the outgoing longwave radiation and $T_s$ is the surface temperature. We use the Buckingham $\Pi$ theorem to derive an analytical function for the IHZ in terms of eccentricity and $\Pi$. We then build a time-dependent energy balance model to resolve the surface temperature evolution and constrain our analytical result. We find that $\Pi$ must be greater than about $\sim 1$ for the mean-stellar flux approximation to be nearly exact and less than about $\sim 0.01$ for the maximum-stellar flux approximation to be nearly exact. In addition to assuming a constant heat capacity, we also consider the effective heat capacity including latent heat (evaporation and precipitation). We find that for planets with an Earth-like ocean, the IHZ should follow the mean-stellar flux limit for all eccentricities. This work will aid in the prioritization of potentially habitable exoplanets with non-zero eccentricity for follow-up characterization.

\end{abstract}

\section{Introduction}\label{sec:intro}

More than 5000 exoplanets have been confirmed to date\footnote{\url{https://exoplanetarchive.ipac.caltech.edu/}}. Prioritization of potentially habitable targets is being carried out using the Habitable Zone (HZ) concept  \citep{2019LUVOIR,2020habex,2021pdaa.book......,2021Hinkel}. The HZ is usually defined as the region around a central star in which an Earth-like planet can potentially have liquid water on its surface \citep{kasting_habitable_1993}. The inner edge of the HZ (IHZ) is determined by the moist greenhouse (or water-loss) limit, where the water vapor content in the stratosphere increases dramatically and then is irreversibly lost to space as a result of photolysis and subsequent hydrogen escape, or by the runaway greenhouse limit, where the atmosphere becomes optically thick with water vapor, limiting the amount of thermal radiation that can be emitted to space, and the oceans evaporate completely \citep{kasting_habitable_1993}. The HZ developed under the assumption of a circular orbit is determined by only one parameter: the effective stellar flux. This study aims to determine a simple approximation of the IHZ for eccentric planets involving the eccentricity and any other necessary parameters.

\begin{figure}
    \centering
    \includegraphics[scale=0.5]{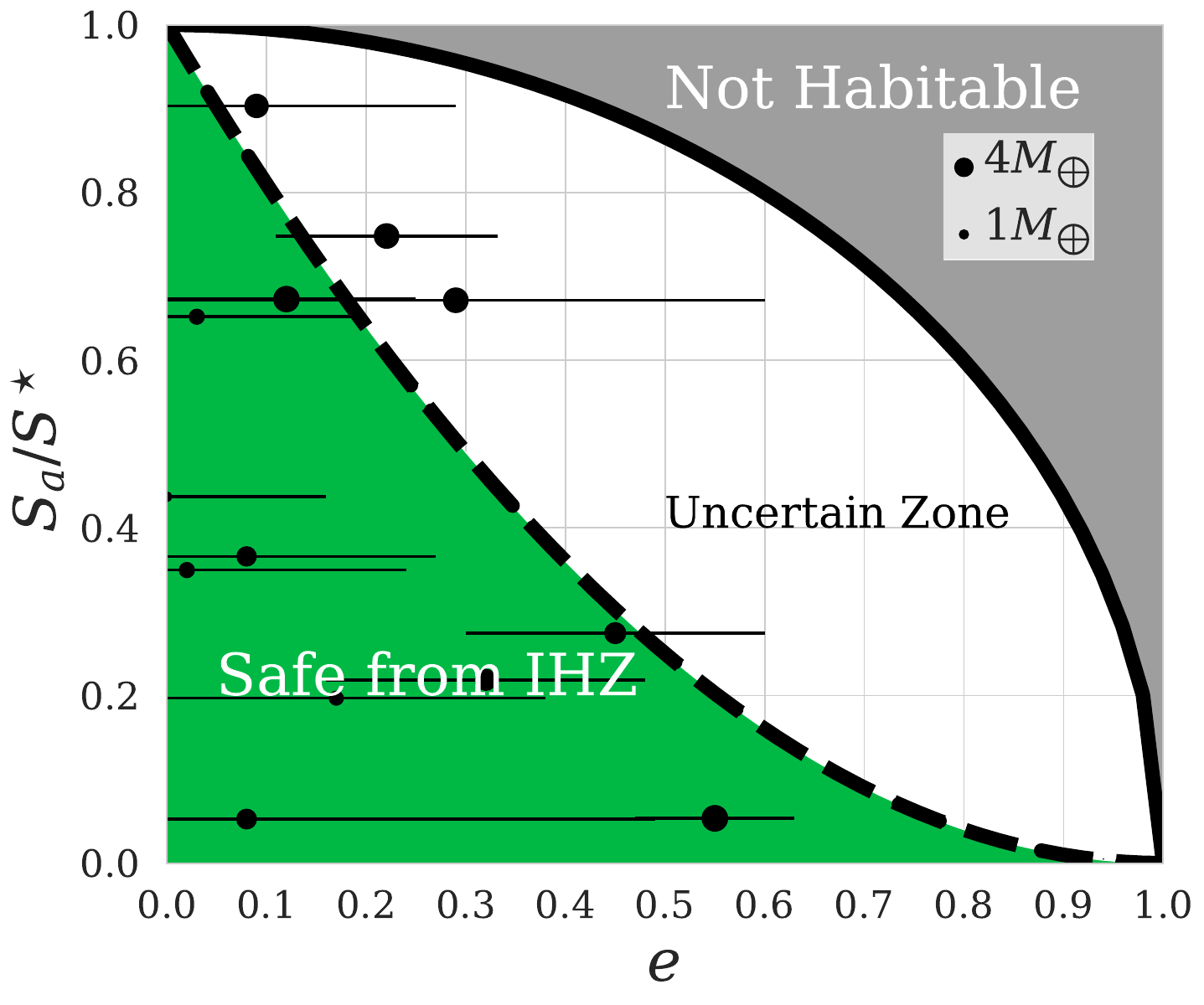}
    \caption{Stellar flux and eccentricity of observed exoplanets with masses $M<8 M_\earth$ \citep{https://doi.org/10.26133/nea12}. $S_a$ is the stellar flux at the semimajor axis of the orbit and $S^\star$ is the stellar flux of the IHZ for a circular orbit, for which we use the moist greenhouse limit from \citet{kopparapu_habitable_2013}. The thick solid black curve represents the mean-stellar flux limit on the IHZ. If $S_a/S^\star$ planet is above this curve a planet is not habitable (Not Habitable). The dashed black curve represents the maximum-stellar flux limit on the IHZ. If $S_a/S^\star$ is lower than this limit, a planet will not experience a moist or runaway greenhouse (Safe from IHZ). For planets in the region in between the limits (Uncertain Zone), climate calculations involving the time evolution of the surface temperature are necessary to determine planetary habitability.  The horizontal bars represent the uncertainties of the observed eccentricities.}
    \label{fig:limit}
\end{figure}

Exoplanet eccentricity can span a considerable range (Fig. \ref{fig:limit}). Eccentricity has been shown to significantly affect seasonal cycles and limits of habitability using climate models of varying complexity \citep{williams_earth-like_2002,dressing_habitable_2010,spiegel_generalized_2010,kadoya_tajika_watanabe_2012,2012Kane,linsenmeier_climate_2015,M_ndez_2017,2017Way,Ohno_2019,2020dynamics,2021Kane}. The heat capacity of the land, oceans, and atmosphere creates thermal inertia in the climate system. Due to this climate inertia, planets can remain habitable even when their instantaneous stellar flux exceeds the critical IHZ value for circular orbits \citep{williams_earth-like_2002,2020Palubski}. However, both \citet{williams_earth-like_2002} and \citet{2020Palubski} considered an Earth-like climate system with a high heat capacity, equivalent to a $\sim$ 50-m column of water. We will show that the behaviors are different for planets with lower heat capacities, and develop a theory for the IHZ involving both eccentricity and a non-dimensional parameter proportional to the heat capacity.

In this paper we consider the effect of eccentricity on the IHZ using theory and a global-mean energy balance model (EBM). We first use the Buckingham $\Pi$ theorem and a consideration of physical limits to derive a theoretical function describing the IHZ as a function of stellar flux, eccentricity, heat capacity, orbital period, and the sensitivity of outgoing longwave radiation to space to surface temperature (section~\ref{sec:theory}). We then constrain the parameters of this function using numerical integration of the EBM and explore the implications of our results in section~\ref{sec:model}. We discuss our results in section~\ref{sec: discussion} and conclude in section~\ref{sec:conclusion}.

\section{Theory}\label{sec:theory}

\subsection{Applying Buckingham Π}\label{sec:buckingham}

We seek the IHZ for an orbit of arbitrary eccentricity. We will specify the IHZ using the stellar flux at the semimajor axis of the orbit, $S_a^\star$, where the superscript $\star$ means that the orbit reaches IHZ. This clearly depends on the eccentricity, $e$, as well as the critical stellar flux of the IHZ for a circular orbit, $S^\star$. We also expect it to depend on variations in surface temperature throughout the orbit, since larger variations in surface temperature make a planet more likely pass out of the HZ. Variations in surface temperature should clearly be influenced by the heat capacity, $C$, and the orbital period, $P$. Additionally they should be influenced by how efficiently radiation to space can damp variations in surface temperature, which is defined by the parameter $B=\frac{\partial \Omega}{\partial T_s}$, where $\Omega$ is the outgoing longwave radiation to space of the planet and $T_s$ is the surface temperature. We will approximate $B$ as a constant in this theory section, which is a good approximation over the entire range of temperatures experienced by modern Earth due to the water vapor feedback \citep{koll_earths_2018} and is sufficient for our theoretical investigations. We will also take the planetary albedo, $\alpha$, to be constant throughout the orbit for simplicity.

We therefore have six physical parameters ($S_a$, $S^\star$, $e$, $C$, $P$, and $B=\frac{\partial \Omega}{\partial T_s}$) and three dimensions (mass, time, and temperature). According to the Buckingham $\Pi$ theorem \citep{PhysRevBuckingham} we can construct a set of three dimensionless parameters to describe the system, and we choose 
\begin{equation}
\begin{split}
    &\Pi_1 = \frac{S_a^\star}{S^\star},\\
    &\Pi_2 = e,\\
    &\Pi\ = \frac{C}{BP}.\\
    \label{eq:pi}
\end{split}
\end{equation}
As a result we can specify stellar flux of the IHZ as follows
\begin{equation}
    \frac{S_a^\star}{S^\star} = F\left(e,\Pi \right),
\end{equation}
where F is a function to be determined.

\subsection{Applying Physical Constraints}\label{sec:PhysicalConstraints}

We can now use physical reasoning to constrain the function $F\left(e,\Pi \right)$.  Consider the two limits where the heat capacity is either infinite or negligible:

\textit{(a) Mean-stellar flux limit}: As the heat capacity approaches infinity, which corresponds to an infinite $\Pi$, the amplitude of surface temperature oscillations will be negligible. The system will therefore remain in the equilibrium state corresponding to the time-averaged energy input and the IHZ will correspond to the orbital mean stellar flux ($\frac{S_a^\star}{\sqrt{1-e^2}}$) exceeding the the critical stellar flux of the IHZ for a circular orbit ($S^\star$), or
\begin{equation}
    \frac{S_a^\star}{S^\star} = \sqrt{1-e^2},
    \label{eq:infinit_Cp}
\end{equation}
If $S_a/S^\star$ is higher than the mean-stellar flux limit, the planet cannot be habitable. 

\textit{(b) Maximum-stellar flux limit}: If the effective heat capacity is negligible ($\Pi \rightarrow 0$), the surface temperature and therefore the outgoing longwave radiation will adjust instantaneously to balance the evolving stellar flux. In this work we will adopt the conservative assumption that the planet is no longer habitable if its maximum surface temperature exceeds the critical circular-IHZ value. Further work would be needed to study the possibility of planets exceeding the critical temperature for some time during their orbit but remaining habitable. We can therefore determine $S_a^\star$ by setting the stellar flux at periastron equal to the critical stellar flux threshold: 
\begin{equation}
    \frac{S_a^\star}{S^\star} = (1-e)^2.
    \label{eq:infinitesimal_Cp}
\end{equation}
This limit is calculated by assuming the largest possible variation in surface temperature, so if $S_a/S^\star$ is lower than this limit, the planet cannot trigger the moist or runaway greenhouse.

\subsection{Ansatz for Function}\label{sec:ansatz}

According to the two IHZ limits outlined in section~\ref{sec:PhysicalConstraints}, we make the following ansatz for the form of the function $F\left(e,\Pi \right)$: 
\begin{equation}
    \frac{S_a^\star}{S^\star} = (1-e)^2+\left(\sqrt{1-e^2}-(1-e)^2 \right) g\left(\Pi \right),
    \label{eq:g_pi}
\end{equation}
where
\begin{equation}
\begin{split}
    &\lim_{\Pi \rightarrow 0} g(\Pi) = 0,\\
    &\lim_{\Pi \rightarrow \infty} g(\Pi) = 1.
    \label{eq:conditions}
\end{split}
\end{equation}

As we expect $\Pi$ to vary over many orders of magnitude, an appropriate ansatz for $g(\Pi)$ yields
\begin{equation}
    \frac{S_a^\star}{S^\star} = (1-e)^2+\left(\sqrt{1-e^2}-(1-e)^2 \right) \frac{1}{2} \left( 1 + \tanh \left(  \frac{\log(\frac{\Pi}{\Pi_0})}{\Delta}  \right) \right),
    \label{eq:g1}
\end{equation}
where $\Pi_0$ is the location of the transition between the two limits and $\Delta$ is the width of the transition in log space. We will use numerical integrations to investigate this ansatz and constrain $\Pi_0 \approx 0.4$ and $\Delta \approx 0.6$ in what follows.
\section{Numerical Results}\label{sec:model}

\subsection{Energy Balance Model}\label{sec:EBM}

We use a time-dependent global-mean energy balance model that describes changes in the global mean surface temperature ($T_s$) forced by periodic variation in incident stellar flux. Considering the energy budget of the planet, any change in the internal energy of the climate system must balance net heating, which is the difference between the net influx of stellar radiation and the outflow of planetary longwave radiation from the top of the atmosphere, 
\begin{equation}
    C \cdot \frac{dT_s}{dt} = \frac{1-\alpha(T_s)}{4}S(e,t) - \Omega(T_s),
\label{eq:EBM}
\end{equation}
where $T_s$ is the surface temperature and $S(e,t)$ is the instantaneous stellar radiation flux at the top of the atmosphere. We assume negligible geothermal heating.

The stellar flux can be scaled with $S_a$ as, 
\begin{equation}
    S(e,t) =  \left( \frac{L_*}{L_\odot}\frac{a_\earth^2}{a^2}\right) \frac{S_\odot}{r^2(e,t)} = \frac{S_a}{r^2(e,t)}
    \label{eq:Sa}
\end{equation}
where $L_*$ and $L_\odot$ are the luminosity of the host star and the Sun, respectively; $a$ and $a_\earth$ are the semi-major axis of the exoplanet and Earth, respectively; $S_\odot$ is the solar constant, which is the average stellar flux that Earth receives; and $r(t)$ is the time-dependent distance from the planet to the star normalized by the semi-major axis, which is a dimensionless quantity that is only dependent on eccentricity and the true anomaly. We calculate $r(t)$ by solving Kepler's equation. 

The criterion for the IHZ is usually defined as the critical stellar flux that can trigger the moist or runaway greenhouse (we will refer to either of these as the extreme climate state). It is obtained using the outgoing longwave radiation and the albedo corresponding to the extreme climate state to calculate the effective stellar flux that can maintain it \citep{kopparapu_habitable_2013}. Denote the critical outgoing longwave radiation associated with the moist or runaway greenhouse as $\Omega ^\star(T^\star)$ where $T^\star$ is the critical temperature. We assume the outgoing longwave radiation and albedo adjust instantaneously with surface temperature, and define the inner edge of the habitable zone (IHZ) as the orbit where the maximum surface temperature of a planet (over an annual cycle) is equal to $T^\star$. $S^\star$ as the critical stellar flux that balances the critical OLR,
\begin{equation}
    S^\star = \frac{4}{1-\alpha(T^\star)}\Omega^\star(T^\star),
    \label{eq:S_ct}
\end{equation}
or equivalently, the stellar flux of the IHZ for a circular orbit.

\begin{figure}
    \centering
    \includegraphics[scale=0.5]{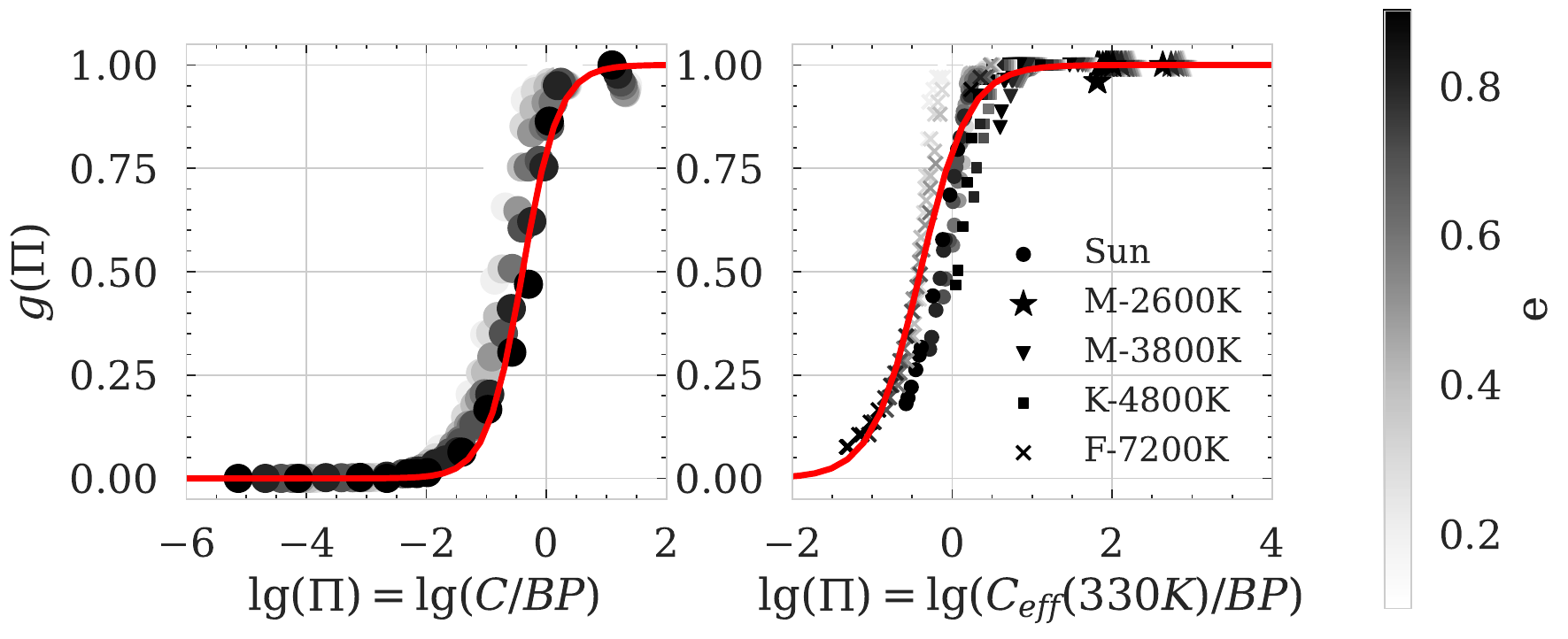}
    \caption{The function defining the transition between the mean-stellar flux limit and the maximum-stellar flux limit ($g(\Pi$), Eq. (\ref{eq:g1}) based on numerical integrations of the EBM assuming a constant heat capacity (left) and a variable heat capacity evaluated at 330 K (Eq. (\ref{eq:cp_complex})) and different stellar types, which affects the orbital period (right). The functional form given in Eq. (\ref{eq:g1}) fits the transition well with $\Pi_0=0.4$ and $\Delta=0.6$.}
    \label{fig:g_pi}
\end{figure}

We integrate the EBM (Eq. \ref{eq:EBM}) to obtain the surface temperature time series and adjust $S_a$ to make the maximum surface temperature during an orbit equal to the critical temperature to determine the IHZ (at which a moist greenhouse or runaway greenhouse occurs). For our numerical integrations we use the OLR ($\Omega(T_s)$) and albedo ($\alpha(T_s)$) functions calculated by \citet{kopparapu_habitable_2013} assuming a saturated radiative-convective atmosphere. We also choose the moist greenhouse limit from \citet{kopparapu_habitable_2013}, of which the critical temperature is $340$K and the critical stellar flux is $S^\star = 1.015 S_0$. This allows us to confirm that Eq.~(\ref{eq:g1}) fits the data well and to constrain $\Pi_0$ to be 0.4 and $\Delta$ to be 0.6 (Fig.~\ref{fig:g_pi}). 

Although our following analysis is based on the results of the circular IHZ of \citet{kopparapu_habitable_2013}, our scaling for the eccentric IHZ applies to any circular-orbit IHZ with the corresponding functions of OLR, albedo, and critical stellar flux. To test the robustness of the $\Pi$ criterion for different circular IHZs, we also extract the functions of OLR and albedo from \citet{Wolf2015} ($B = 1.18  \mathrm{W\cdot K^{-1}\cdot m^{-2}}$), and accept their critical stellar flux and temperature ($S^\star = 1.125 S_0; T_s = 331.9\mathrm{K}$) to perform the same calculation. The results constrain $\Pi_0$ and $\Delta$ to the same values as shown in Fig.~\ref{fig:g_pi}.

\subsection{Variable Heat Capacity} \label{sec:heatcapacity}

\begin{figure}
    \centering
    \includegraphics[scale=0.3]{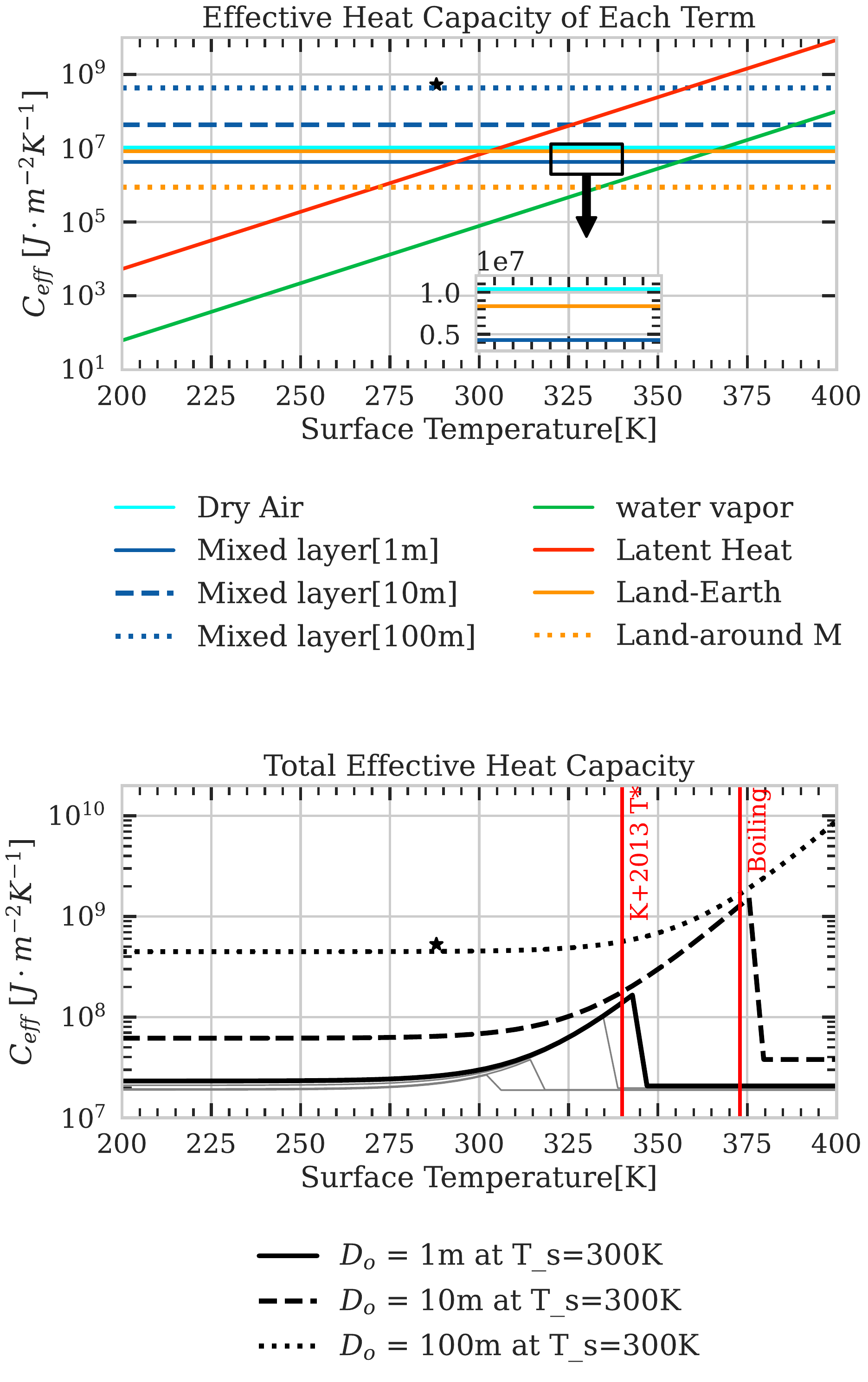}
    \caption{Terms contributing to the effective heat capacity (upper) and the effective heat capacity of planets with different ocean depths (lower) as a function of temperature. Upper: The inset zooms in on the small difference between dry air, land, and a 1-m mixed layer. ``Land" is calculated with modern Earth's orbital period and ``Land around M" is calculated for an equivalent planet with a smaller orbital period orbiting an M-dwarf ($0.09M_\odot$). The black star shows the observational effective global-mean heat capacity of modern Earth considering the measurements of heat content of ocean and the surface temperature changes in decades \citep{schwartz_heat_2007}. Lower: The red vertical lines mark the critical temperature for a moist greenhouse from \citet{kopparapu_habitable_2013} and the boiling point on Earth. The thinner black curves denote the the effective heat capacity of planets with ocean depths of 0.01m, 0.1m, 0.5m from left to right.}
    \label{fig:Cp}
\end{figure}

Up to now we have assumed that the heat capacity of a planet is fixed, but it can actually vary with time as a result of a variety of factors. We will now consider an effective heat capacity ($C_{\textit{eff}}$) including energy storage by land, mixed ocean, dry air, and water vapor in the atmosphere, as well as latent heat from evaporation of water. 

For simplicity, we assume a rapidly mixed one-layer ocean that is uniform in temperature, so the heat capacity of the ocean is simply $C_{p,o} = D_o \cdot \rho_o \cdot c_{p,o}$ where $D_o$ is the depth of the mixed layer, $\rho_o$ is the density of the ocean, and $c_{p,o}$ is the specific heat capacity of ocean water. Absent information on the salinity of the ocean, we assume Earth's values: $\rho_o = 1020 \ \mathrm{kg\cdot m^{-3}}$ and $c_{p,o} = 4200\  \mathrm{J\cdot kg^{-1}  K^{-1}}$ \citep{hartmann_global_1994}. Note that the depth of the mixed layer changes with the surface temperature as a result of the evaporation water. The land as a solid can only transfer heat through conduction and there is a characteristic depth ($h_L$) through which a temperature anomaly on the surface associated with periodic forcing will penetrate in one period \citep{turcotte_geodynamics_2002}. 
\begin{equation}
\begin{split}
    C_{p,L} &= h_L\cdot \rho_L\cdot c_{p,L}\\
    h_L &= \sqrt{\kappa \tau}
\end{split}
\end{equation}
where $\rho_L = 2600 \mathrm{km\cdot m^{-3}}$ is the density of the soil, $c_{p,L} = 733 \mathrm{J\cdot \mathrm{kg^{-1}  K^{-1}}}$ is the specific heat capacity of the soil, $\kappa = 5 \times 10^{-7} \mathrm{m^2 s^{-1}}$ is the thermal diffusivity of the soil \citep{hartmann_global_1994,abbot_importance_2010}, and $\tau$ is the duration of forcing, that is, the orbital period, in this case. 

Assuming an Earth-like atmosphere, the heat capacity of dry air is $C_{p,a} = M_a\cdot c_{p,a}$, where the mass per unit area $M_a$ is the Earth's value and $c_{p,a} = 1004 \mathrm{J\cdot kg^{-1}  K^{-1}}$. The heat capacity of water vapor per unit area is $C_{p,v} = M_v\cdot c_{p,v}$, where the specific heat capacity of water vapor is $c_{p,v} = 1880 \mathrm{J\cdot kg^{-1}  K^{-1}}$ (at 350K from \citet{noauthor_water_2005}\footnote{Accessed 3 Oct. 2022}). The mass of water vapor per unit area as a function of surface temperature ($M_v$) is obtained from the Extended Data Fig. 3 of \citet{2013Natur.504..268L}. We assume the water vapor mixing ratio is a slave to the surface temperature and adjusts instantaneously. Water vapor in the atmosphere increases with surface temperature as a result of evaporation, leading to the consumption of latent heat. The latent heat of water vaporization per unit area ($L_v$) is given by $L_v = l_v\frac{dM_v(T_s)}{dT_s}$, where the specific latent heat of water vaporization is $l_v = 2.3\times 10^6 \mathrm{J\cdot kg^{-1}}$. Latent heat is the largest contributer to the total $C_{\textit{eff}}$ near the critical temperature, if the global mean depth of the ocean is $\lesssim 10$ m.

Overall, the total effective heat capacity can be expressed as

\begin{equation}                                                                                                                                                                                                                  
    C_{\textit{eff}} = D_o(T_s) \cdot \rho_o \cdot c_{p,o} + C_{p,L} + C_{p,a} + M_v (T_s) \cdot c_{p,v} + l_v\frac{dM_v(T_s)}{dT_s}
    \label{eq:cp_complex}
\end{equation}
where the heat capacity of dry air ($C_{p,a}$) and land ($C_{p,L}$) are constants for a given planet, while the latent heat and the heat capacity of ocean and water vapor change with the surface temperature due to vaporization. We plot the various terms in $C_{\textit{eff}}$ in Fig.~\ref{fig:Cp}. The right panel Fig. \ref{fig:g_pi} shows that if we calculate $C_{\textit{eff}}$ at 330K, our results approximately collapse onto the functional form we found for the fixed heat case (Eq.~(\ref{eq:g1})).

\subsection{Implications of Results}\label{sec:implications}

Eq.~(\ref{eq:g1}) represents a specification of the IHZ for planets with non-zero eccentricity and we have now constrained its parameters, so in a sense our work is done; however, it is worthwhile to consider some of the implications of our results in graphical form.

\begin{figure}
    \centering
    \includegraphics[scale=0.45]{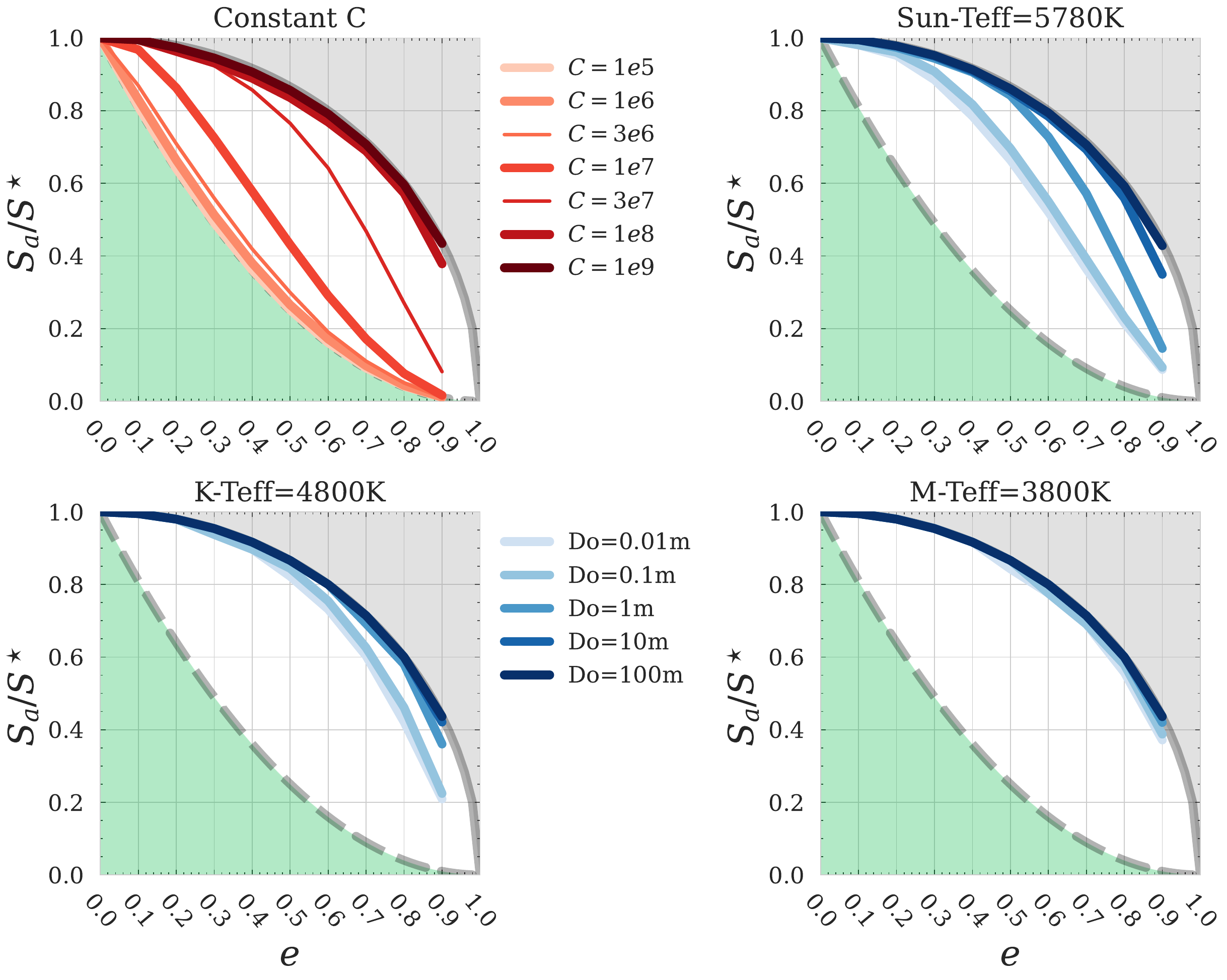}
    \caption{The same layout as Fig. \ref{fig:limit} but with the IHZ calculated assuming different surface conditions for planets in the uncertain zone. Top Left: The effective heat capacity is assumed to be constant. Top Right: The effective heat capacity includes potentially variable terms from land, ocean, dry atmosphere, water vapor and latent heat. Bottom: Lower: The same simulations as shown in the top right panel but for different types of the host stars. $S^\star$ and the function of albedo are adapted according to \citet{kopparapu_habitable_2013}. Assuming a planet with an Earth-like atmosphere, it is always safe to apply the mean-stellar flux limit for M-dwarfs and K-dwarfs because of shorter orbital periods. }
    \label{fig:IHZ_Cp}
\end{figure}

We first show numerical results for the IHZ as a function of stellar flux and eccentricity assuming a constant heat capacity (Fig. \ref{fig:IHZ_Cp}). Planets with $C<10^6 \mathrm{J/m^{2} K}$ follow the maximum-stellar flux limit very well, while planets with $C>10^8 \mathrm{J/m^{2} K}$ follow the mean-stellar flux limit. Interestingly, we expect that many potentially habitable planets have $C \sim 10^7-10^8 \mathrm{J/m^{2} K}$ in the intermediate range between these two limits (Fig.~\ref{fig:Cp}).

We also integrate our EBM with a variable effective heat capacity (Eq. (\ref{eq:cp_complex})) for a variety of ocean depths (Fig.~\ref{fig:IHZ_Cp}). We assume a planet with an Earth-like atmosphere, but we consider the difference in land heat capacity due to changing orbital periods. For lower eccentricities ($e<0.5$), if planets have an ocean deeper than 1~m, the IHZ follows the mean-stellar flux limit. For $e>0.5$, the ocean must be deeper than 10~m for the mean-stear flux limit to apply. For planets with an atmosphere of Earth's mass, even dry air provides a heat capacity of $\sim 2 \times \mathrm{10^7 J/m^{2} K}$. This provides sufficient thermal inertia to keep them far from the maximum-stellar flux limit, even for a negligible ocean (Fig.~\ref{fig:IHZ_Cp}). Note that this conclusion is dependent on the column heat capacity of dry air in a planet's atmosphere.

We also calculate the IHZ for planets orbiting different types of host stars (Fig. \ref{fig:IHZ_Cp}). Changes in the orbital period near the IHZ lead to different behavior even with the same surface conditions through their effect on $\Pi$ (Eq.~(\ref{eq:pi})). As discussed above, however, the results for different stars do approximately collapse onto the same IHZ function (Fig.  \ref{fig:g_pi}). It is interesting to note that planets orbiting M dwarfs have a short enough period that $\Pi$ is large (Eq.~\ref{eq:pi}) and they nearly follow the mean-flux limit even if their only heat capacity is an Earth-like atmosphere with no ocean (Fig. \ref{fig:IHZ_Cp}).

\begin{figure}
    \centering
    \includegraphics[scale = 0.5]{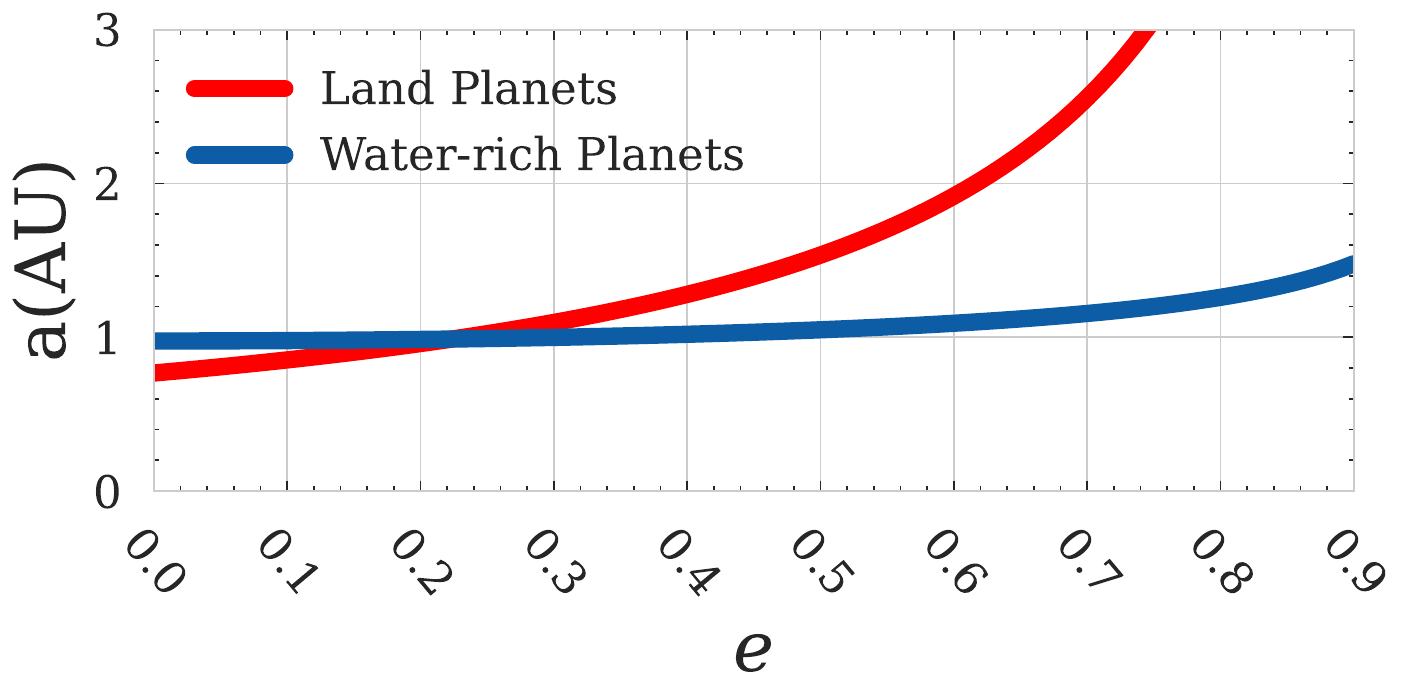}
    \caption{Comparison of the IHZ (given in terms of semi-major axis) as a function of eccentricity for a land planet and a water-rich planet. We assume that the stellar flux of the circular IHZ is 1.7 times that of modern Earth's for the land planet \citep{doi:10.1089/ast.2010.0545} and 1.015 times that of modern Earth's for the water-rich planet \citep{kopparapu_habitable_2013}. We assume that the land planet follows the maximum-stellar flux limit and the water-rich planet follows the mean-stellar flux limit.}
    \label{fig:IHZ_land}
\end{figure}

Previous work has shown that for circular orbits $S^\ast$ should be larger for dry planets than for planets with an ocean \citep{doi:10.1089/ast.2010.0545,2015Kodama}. Our work suggests that if the eccentricity is large enough, $S_a^\ast$ will be larger for ocean planets than a dry planets with a low enough heat capacity. For example, a dry planet with a Mars-like atmosphere and less-conductive soil would have an effective heat capacity of only $\simeq \mathrm{10^6 J/m^{2}} K$, forcing the IHZ to the maximum-stellar flux limit. To illustrate this effect, we show that the semi-major axis corresponding to IHZ for land planets following the maximum-stellar flux limit and for water-rich planets following the mean-stellar flux limit cross at an eccentricity of approximately 0.2 (Fig.~\ref{fig:IHZ_land}).

\section{Discussion}\label{sec: discussion}

This work is a theoretical investigation into the factors that determine the IHZ for eccentric planets. 3D effects such as atmospheric circulation, heat transport, humidity, and clouds may add additional complexities that modify the theory developed here. For example, we assume the water vapor mixing ratio and OLR adjust with the surface temperature instantaneously, which may not be exactly correct \citep{williams_earth-like_2002,2017Way}. Investigation of the level of deviation this causes from the IHZ limit we have derived (Eq.~\ref{eq:g1}) requires detailed 3D modelling. 

Our model simplifies the climate system to a global mean state. This means that it may be less applicable to planets orbiting M-stars, which are likely to be tidally locked. On the other hand, as the IHZ is approached atmospheric heat transport from the day side to the night side of synchronously rotating planets becomes so large that the surface temperature is nearly uniform \citep{2014Yang}. Moreover, planets with non-zero eccentricities may be caught in higher order spin-orbit states, which should lead to more uniform temperature distributions \citep{barnes2017}. Ultimately, this is another effect that should be checked with 3D GCM calculations. 


In this work we focused on the inner edge of the habitable zone and did not consider the outer edge. Most of our numerical simulations with relatively small heat capacities ($C_p \lesssim 3\times 10^7 J\cdot m^{-2}\cdot K^{-1}$) and high eccentricity (e$>$0.4) have  minimum temperatures below 0$^\circ$C. However, this does not necessarily mean that they are beyond the outer edge of the habitable zone, as experience in terrestrial locations such as Chicago demonstrates. Full investigation of these issues is beyond the scope of the current work.

\section{Conclusion}\label{sec:conclusion}

Our main conclusions are as follows:

\begin{itemize}
    \item Using Buckingham $\Pi$ theory, physical arguments, and numerical integrations, we find that the stellar flux at the semimajor axis of the inner edge of the habitable zone (IHZ) for eccentric planets can be approximated as
\begin{equation}
    \frac{S_{a}^\ast}{S^\star} = (1-e)^2+\left(\sqrt{1-e^2}-(1-e)^2 \right) \frac{1}{2} \left( 1 + \tanh \left(  \frac{\log(\frac{\Pi}{\Pi_0})}{\Delta}  \right) \right),
\end{equation}
where $\Pi = \frac{C}{BP}$, where $C$ is the heat capacity of the planet, $P$ is the orbital period,  $B=\frac{\partial \Omega}{\partial T_s}$, where $\Omega$ is the outgoing longwave radiation to space of the planet and $T_s$ is the surface temperature, $\Pi_0 \approx 0.4$, and $\Delta \approx 0.6$. 
    \item To put this in concrete terms, a potentially habitable planet orbiting a G star with an ocean at least 10~m deep should have surface temperature variations small enough that its IHZ is set by the mean-stellar flux it receives ($\frac{S_{a}^\ast}{S^\star}=\sqrt{1-e^2}$).
    \item Moreover, planets with shorter periods (orbiting M or K stars) only require a heat capacity equivalent to Earth's atmosphere (without any ocean) in order to follow the mean-stellar flux limit.
    \item Finally, our calculations suggest that although the IHZ for a dry planet with a circular orbit may occur at a higher stellar flux than a planet with an ocean, this situation reverses for eccentricities above $\approx0.2$ because temperature oscillations on dry planets become large. 
\end{itemize}
 

This research has made use of the NASA Exoplanet Archive, which is operated by the California Institute of Technology, under contract with the National Aeronautics and Space Administration under the Exoplanet Exploration Program. This work was supported by NASA award number 80NSSC21K1718, which is part of the Habitable Worlds program. This work was supported by the NASA Astrobiology Program grant No. 80NSSC18K0829 and benefited from participation in the NASA Nexus for Exoplanet Systems Science research coordination network. We acknowledge the funding support by the NASA Exoplanet Research Program (NNH18ZDA001N-2XRP), which was conducted at the Jet propulsion Laboratory, California Institute of Technology, under contract by NASA. This work was completed with resources provided by the University of Chicago Research Computing Center. We thank Stephen Kane, Huanzhou yang, Bowen Fan and Michael Way for helpful discussions and feedback. We also thank the  the anonymous reviewer for their comments which have improved the manuscript.

\software{Astropy \citep{astropy:2013, astropy:2018, astropy:2022}, iPython \citep{PER-GRA:2007},  Jupyter \citep{Kluyver2016jupyter}, Matplotlib \citep{Hunter:2007}, Science Plots \citep{SciencePlots}, Scipy \citep{2020SciPy-NMeth}, Seaborn \citep{Waskom2021}, Pandas \citep{reback2020pandas}}


\bibliography{eccentric_IHZ.bib}
\bibliographystyle{aasjournal}

\end{CJK*}
\end{document}